\documentclass{acm_proc_article-sp}
\usepackage{graphicx}
\usepackage{balance} 
\usepackage{multirow}
\usepackage{listings}
\usepackage{url}
\usepackage[usenames,dvipsnames]{color}
\begin{document}

\title{Cell Stores}

\numberofauthors{1}

\author{
\alignauthor
Ghislain Fourny\\
       \affaddr{28msec, Inc.}\\
       \affaddr{Z\"urich, Switzerland}\\
       \email{g@28.io}
}
\date{3 October 2014}
\maketitle

\begin{abstract}
Cell stores provide a relational-like, tabular level of abstraction to business users while leveraging recent database technologies, such as key-value stores and document stores. This allows to scale up and out the efficient storage and retrieval of highly dimensional data. Cells are the primary citizens and exist in different forms, which can be explained with an analogy to the state of matter: as a gas for efficient storage, as a solid for efficient retrieval, and as a liquid for efficient interaction with the business users. Cell stores were abstracted from, and are compatible with the XBRL standard for importing and exporting data. The first cell store repository contains roughly 200GB of SEC filings data, and proves that retrieving data cubes can be performed in real time (the threshold acceptable by a human user being at most a few seconds).
\end{abstract}

\section{Introduction}
In 1970, Codd \cite{Codd1970} introduced the relational model as an alternative to the graph and network models (such as file systems) in order to provide a more suitable interface to users, and to protect them from internal representations (``data independence'').

The relational model's first implementation was made public in 1976 by IBM \cite{Astrahan1976}.

In the last four decades, the relational model has been enjoying undisputed popularity and has been widely used in enterprise environments. This is probably because it is both very simple to understand and universal. Furthermore, it is accessible to business users without IT knowledge, to whom tabular structures are very natural --- as demonstrated by the strong usage of spreadsheet software \cite{Mattessich1958} \cite{Mattessich1961} (such as Microsoft Excel, Apple Numbers, Lotus 1-2-3, OpenOffice Calc) as well as user-friendly front-ends (such as Microsoft Access).

However, in the years 2000s, the exponential explosion of the quantity of data to deal with increasingly showed the limitations of this model. Several companies, such as Google, Facebook or Twitter needed scaling up and out beyond the capabilities of any RDBMS, both because of the \emph{quantity} of data (rows), and because of the \emph{high dimensionality} of this data (columns). Each of them built their own, ad-hoc data management system (Big Table \cite{Chang2008}, Cassandra \cite{Lakshman2010}, ...). These technologies often share the same design ideas (scale out through clustering and replication, high dimensionality handling through data heterogeneity and tree structures), which led to the popular common denomination of NoSQL, a common roof for:
\begin{description}
\item[Key-value stores,] which store big collections of key-value pairs. Example: DynamoDB.
\item[Document stores,] which are document-oriented, typically supporting XML \cite{XML} or JSON \cite{JSON}. Example: MongoDB.
\item[Column stores,] which keep the table abstraction while allowing some sparseness. Example: Cassandra.
\item[Graph databases,] which work at the lower triple level. Example: Neo4j.
\end{description}

NoSQL solves the scale-up issue, but at a two-fold cost:
\begin{description}
\item[For developers,] the level of abstraction provided by NoSQL stores is much lower than that of the relational model. These data stores often provide limited querying capability such as point or range queries, insert, delete and update (CRUD). Higher-level operations such as joins must be implemented in a host language, on a case by case basis.
\item[For business users,] these data models are much less natural than tabular data. Reading and editing data formats such as XML, JSON requires at least basic IT knowledge. Furthermore, business users should not have to deal with indexes at all.
\end{description}

This is a major step back from Codd's intentions back in the 1970s, as the very representations he wanted to protect users from (tree-like data structures, storage, ...) are pushed back to the user.

Reluctance can be observed amongst non-technical users, and this might explain why the ``big three'' (Oracle, Microsoft, IBM) are heavily pushing towards using of the SQL language \cite{Chamberlin1974} on top of these data stores.

This paper introduces the cell store data paradigm, whose goal is to (i) leverage the technological advancements made in the last decade, while (ii) bringing back to business users control and understanding over their data. The cell store paradigm was vastly inspired by and abstracted from the XBRL standard \cite{XBRL}, which defines a serialization format for exchanging facts. Historically, cell stores were precisely designed in order to efficiently store and retrieve XBRL data. With time, this paradigm was decoupled from XBRL in such a way that it could also accommodate for data beyond business reporting. In particular, relational data can also be dropped into a cell store.

Cell stores are at a sweet spot between on the one hand key-value stores, in that they scale up seamlessly and gracefully in the quantity of data as well as the dimensionality of the data, and on the other hand the relational model, in that business users access the data in tabular views via familiar, spreadsheet-like interfaces.

Section \ref{section-state-of-the-art} gives an overview of state of the art technologies for storing large quantities of highly dimensional data and their shortcomings. Section \ref{section-why} motivates the need for the cell stores paradigm. Section \ref{section-data-model} introduces the data model behind cell stores. Section \ref{section-relational-mapping} shows how a relational database can be stored naturally in a cell store. Section \ref{section-xbrl-standard} points out that there is a standard format, XBRL, for exchanging data between cell stores as well as other databases. Section \ref{section-implementation} gives implementation-level details. Section \ref{section-performance} explores performance.

\section{State of the art}
Before introducing the cell store paradigm in details, we quickly survey the current database/datastore landscape.

\label{section-state-of-the-art}

\subsection{Relational databases}

Relational databases are a very mature and stable technology, used everywhere in the world. It is based on the entity-relationship model and the powerful relational algebra, relying on the functional and declarative SQL language \cite{Chamberlin1974}. It has the advantage that tables are very business friendly and easy to understand.

However, relational databases showed their limits in the last decade, because they are monolithic and hard to scale up and out when the amount of data reaches the Terabyte to Petabyte range. It is very hard and expensive to make a relational schema evolve when the data is spread across multiple machines.

Also, it is very challenging to maintain ACID properties \cite{Haerder1983} beyond one machine, which is why a newer generation of databases was designed, dubbed as NoSQL even though they are diverse (key-value stores, document stores, column stores, etc.). ACID got replaced with the idea induced by the CAP theorem \cite{Gilbert2002} that consistency must be relaxed in order to ensure availability and partition tolerance.

\subsection{Key-value stores}

Key-value stores provide a very simple level of abstraction, organizing the data as collections of key-value pairs. A collection can be partitioned across several machines, as well as replicated. Indexes allow very efficient data retrieval. Key-value stores are very friendly to powerful parallelism frameworks such as MapReduce, as stateless mappings can be performed in parallel in each location where the data lies.

Key-value stores offer a very low-level interface that requires programmatic abilities to interact with.

An example of popular key-value store is Amazon DynamoDB \cite{DeCandia2007}.

\subsection{Document stores}

Document stores are centered on the concept of document. It can be seen as a key-value store where values are not black boxes, but instead are XML \cite{XML}, JSON \cite{JSON}, YAML \cite{YAML} or protocol buffers \cite{ProtocolBuffers} (or even word processor, spreadsheet, files, ...) documents. Documents are often arborescent and organized in heterogeneous collections. Secondary indexes can be built based on the content of these documents.

Document stores often offer a very basic query language that allows filtering and projecting documents. They require a host language on top to implement more elaborate functionality such as joins. 

Popular document stores include MongoDB \cite{MongoDB}, Cloudant \cite{Cloudant}, CouchDB \cite{CouchDB}, ElasticSearch \cite{ElasticSearch}.

\subsection{Column stores}

Column stores, like relational databases, are table centric, but offer much more flexibility. In particular, they can be very sparse, because each row (specifically, a sequence of columns) may have several absent columns (heterogeneity). Column store typically denormalize the data, optimizing projection and selection, avoiding the need for joins as much as possible.

However, private key columns are rigid, and all rows within a table must have exactly the private keys required by the schema. Tables must be created for each different private key topology.

Popular column stores include BigTable \cite{Chang2008}, HBase \cite{HBase}, Cassandra \cite{Lakshman2010}.

\subsection{OLAP}

OLAP \cite{Codd1993} stands for OnLine Analytical Processing and targets dimensional data (data cubes). Data can be sliced and diced, aggregated (roll ups). OLAP is compatible with spreadsheet front ends using pivot table to visualize the data in a business-friendly way. The MDX language \cite{Nolan1999} is a standard way of querying for data cubes. There are two main flavors of OLAP:
\begin{description}
\item[ROLAP] Data cubes are stored in tables organized in a star or snowflake setting: a central table with the data, and one additional table for each dimension. ROLAP is very rigid, and tables must be created for each data cube.
\item[MOLAP] Data cubes are stored in an efficient proprietary format in memory. Data cubes that are queried often are precomputed and pre-aggregated. MOLAP reaches its limits as soon as data cube queries are more diverse and hard to predict in advance.
\end{description}

A third flavor, HOLAP, is a hybrid of the two. OLAP does not scale up well beyond a few hundred dimensions.

The main vendors (IBM, Microsoft, Oracle, SAP) offer an OLAP implementation.

\subsection{Graph databases}

Graph databases manipulate graphs, mostly implemented as collections of triples (subject, attribute, object). The most used query language is SPARQL \cite{SPARQL}.

Graph databases are very useful when dealing with semantic data, ontologies and AI. However, when dealing with structured data, they become inefficient, because each single structured query needs to join multiple triples and aggregate them back into a meaningful format.

Graph database implementations include ArangoDB \cite{ArangoDB}, Neo4j \cite{Neo4j}.

\subsection{Spreadsheet software}

Surprisingly, the biggest database in the world might well be all those spreadsheet files lying around in mail boxes. This illustrates an impedance mismatch between business use cases and database solutions.

Creating a database or a table on the servers often requires interacting with IT administrators. Many business users end up filling in their data into a spreadsheet and sending it to their colleagues. The data is copied -- sometimes even rekeyed from printed paper -- and sent again. This leads to:

\begin{description}
\item[Data duplication] There exists several versions of the same data.
\item [Inconsistencies] It is not clear where the latest data is, and people might not agree or not know which values are correct among multiple files.
\item[Mistakes] Upon copying or rekeying, mistakes can be introduced by humans that could have been avoided with a database.
\item[High HR costs] People copying, rekeying and sending e-mails has a concrete cost.
\item[Information leak] E-mails can be sent, by mistake or not, outside of the company.
\end{description}

Cell stores aim at keeping the excellent and proven spreadsheet interface, while fixing these issues by seamlessly integrating the spreadsheet with a database backend.

\section{Why cell stores}
\label{section-why}

Cell stores leverage the advantage of the aforementioned state-of-the-art technologies:

\begin{itemize}
\item Like key-value stores and document stores, they scale out with heterogeneous data. The data can be distributed across a cluster, replicated, and efficiently retrieved. They are also compatible with MapReduce-like parallelism paradigms.
\item Like the relational model, cell stores expose the table abstraction.
\item Like column stores, cell stores focus on projection and selection, and denormalize the data. 
\item Like document stores, schemas are not needed upfront and can be provided at will at query time.
\item Like OLAP, cell stores expose the data cube abstraction to the user.
\item Cell stores can handle highly dimensional data, because storage works at the cell level.
\item Cell stores expose the data via a familiar spreadsheet-like interface to the business users, who are in complete control of their taxonomies (schemas) and rules.
\end{itemize}

\section{The Cell Store Data Model}
\label{section-data-model}

We now introduce the data model behind cell stores. All examples are fictitious (names, numbers).

\subsection{Gas of cells}

As the name ``cell store'' indicates, the first class citizen in this paradigm is the cell. In the OLAP and XBRL \cite{XBRL} universes, it is also called fact, or measure. In the spreadsheet universe, it really corresponds to a cell. In the relational database universe, it corresponds to a single value on a row.

The cell can be seen as an atom of data, in that it represents the smallest possibly reportable unit of data. It has a single value, and this value is associated with dimensional coordinates that are string-value pairs. These dimensional coordinates are also called aspects, or properties, or characteristics. They uniquely identify a cell, and a consistent cell store should not contain any two cells with the exact same dimensional pairs. Nevertheless, cell stores are able to handle collisions elegantly, that is, no fatal error is thrown if or when this happens.

There are no limits to the number of dimension names and their value space. Cell stores scale up seamlessly with the total number of dimensions. There is only one required dimension called \emph{concept}, which describes \emph{what} the value represents. All other dimensions are left to the user's imagination, although typically a validity period (instant or duration: when), an entity (who), a unit (of what), a transaction time, etc, are to be commonly found as well.

Figure \ref{fig-cell} shows an example of a single cell.

\begin{figure}
\centering
\caption{A cell. Each dimension is associated with a value. The value of the cell is shown at the bottom.}
\label{fig-cell}
\vspace{3mm}
\begin{tabular}{|l|l|}
\hline
Dimension & Value \\
\hline
Concept & Assets \\
Period & Sept. 30th, 2012 \\
Entity & Visto \\
Unit & US Dollars \\
Region & United States \\
\hline
\multicolumn{2}{|c|}{\textbf{3,000,000,000}} \\
\hline
\end{tabular}
\end{figure}

The main idea of the cell store paradigm is that there is a single one, big collection of cells. All the data is in this collection, and on a logical level, this collection is not partitioned or ordered in any (logical) way. An analogy can be made with a gas of molecules, where the molecules fly around without any particular order or structure.

In the same way as gas can be stored in containers, the cell gas can (should) be clustered and replicated to enhance the performance of the cell store. Whether clustering is done randomly or following a pattern based on dimension values is mostly driven by optimization and performance based on the use case.

\subsection{Hypercubes}

After cells, the next most important construct is the hypercube. In the cell store paradigm, hypercubes are queries that correspond to both selection and projection in the relational algebra. Unlike in the relational algebra though, selection and projection are the same.

\subsubsection{Point queries and indices}

Now that we have a gas of cells available, we can begin to play with it. The first idea that comes to mind is how to retrieve a cell from the gas (point query).

Point queries leverage the index capabilities of the underlying storage layer. If the cell gas is small and contains many concepts, a single hash index on the \emph{concept} dimension will be enough. For bigger cell gases, other techniques allow scaling up, such as:
\begin{itemize}
\item compound keys: a single index on several dimensions such as \emph{concept}, \emph{period} and \emph{entity}.
\item separate hash keys: use single indices separately, and compute their intersection.
\end{itemize}

\subsubsection{Hypercube queries}
In technologies such as OLAP, the first class citizen is the hypercube, which can be seen as the \emph{schema}. In cell stores, the hypercube can be seen as the \emph{query}.

A hypercube is a dimensional range (as opposed to dimensional coordinates). It is made of a set of dimensions, and each dimension is associated with a range, which is a set of values. The range can be either an explicit enumeration (for example, for strings), or an interval (like the integers between 10 and 20), or also more complex multi-dimensional ranges (consider Geographic Information Systems (GIS)).

Figure \ref{fig-hypercube} shows an example of hypercube. It looks a bit like a cell, except that there is no value, and dimensions are associated with ranges rather than single values.

A cell belongs to a hypercube if:

\begin{itemize}
\item it has exactly the same dimensions
\item for each dimension, the value belongs to the domain of that dimension as specified in the hypercube
\end{itemize}

Hypercubes may (and will typically) have missing cells or even be sparse.

Figure \ref{fig-hypercube} also shows two cells satisfying the above criterion.

\begin{figure}
\centering
\caption{A hypercube containing 18 cells. Each dimension is associated with a range or set of values. Below it, two cells within this hypercube are shown.}
\label{fig-hypercube}
\vspace{3mm}
\begin{tabular}{|l|l|}
\hline
Dimension & Value \\
\hline
Concept & Assets, Equity, Liabilities \\
Period & Sept. 30th, 2012, Dec. 31st, 2012 \\
Entity & Visto, Championcard, American Rapid \\
Unit & US Dollars \\
\hline
\end{tabular}

\begin{tabular}{|l|l|}
\hline
Dimension & Value \\
\hline
Concept & Equity \\
Period & Dec. 31st, 2012 \\
Entity & Championcard \\
Unit & US Dollars \\
\hline
\multicolumn{2}{|c|}{\textbf{5,000,000,000}} \\
\hline
\end{tabular}
\begin{tabular}{|l|l|}
\hline
Dimension & Value \\
\hline
Concept & Liabilities \\
Period & Dec. 31st, 2012 \\
Entity & American Rapid \\
Unit & US Dollars \\
\hline
\multicolumn{2}{|c|}{\textbf{3,000,000,000}} \\
\hline
\end{tabular}
\end{figure}

Like point queries, hypercube queries also leverage indices. Range indices, in addition to, or as an alternative to hash indices, prove particularly useful in the case of numeric or date dimension values. Domain-specific indices like GIS also fit well in this picture.

\subsubsection{Default dimension values}

In cell stores, the number of dimensions and their names vary across cells. Hypercube queries accommodate for this flexibility with the notion of a default dimension value.

Figure \ref{fig-default} shows a hypercube that defines a default value of ``[World]'' for the ``Region'' dimension.

If a hypercube specifies a default value for a given dimension, then the condition that a cell must have that dimension to be included in the hypercube is relaxed. In particular, a cell will also be included if it does not have a ``Region'' dimension. When this happens, an additional dimensional pair is added to the cell on the fly, using the default value as value. This implies that in the end, the set of cells that gets returned for the hypercube query always has exactly the dimensions specified in the hypercube.

\begin{figure}
\centering
\caption{A hypercube using a default dimension value (shown in square brackets). Below it, two cells within this hypercube are shown. In the second cell, the default value was automatically inserted, although it does not appear in the original cell.}
\vspace{3mm}
\label{fig-default}
\begin{tabular}{|l|l|}
\hline
Dimension & Value \\
\hline
Concept & Assets, Equity, Liabilities \\
Period & Sept. 30th, 2012 \\
Entity & Visto, Championcard, American Rapid \\
Unit & US Dollars \\
Region & United States, [World] \\
\hline
\end{tabular}

\begin{tabular}{|l|l|}
\hline
Dimension & Value \\
\hline
Concept & Assets \\
Period & Sept. 30th, 2012 \\
Entity & Visto \\
Unit & US Dollars \\
Region & United States \\
\hline
\multicolumn{2}{|c|}{\textbf{3,000,000,000}} \\
\hline
\end{tabular}

\begin{tabular}{|l|l|}
\hline
Dimension & Value \\
\hline
Concept & Assets \\
Period & Sept. 30th, 2012 \\
Entity & Visto \\
Unit & US Dollars \\
\emph{Region} & \emph{[World]} \\
\hline
\multicolumn{2}{|c|}{\textbf{4,000,000,000}} \\
\hline
\end{tabular}
\end{figure}

In particular, a hypercube is highly structured.

\subsubsection{The ``Big Cube''}
Theoretically, it would be feasible to build a hypercube with all dimensions used in the gas of cells, allowing default values for all of these. Then all cells would belong to this hypercube. However, this is an extremely sparse hypercube, and the size of this hypercube would typically be orders of magnitude greater than the entire visible universe.

\subsubsection{Materialized hypercube}

The answer to a hypercube query can be showed in a consolidated way, resembling a relational table. Each column corresponds to a dimension, and the last column to the value. Figure \ref{fig-materialized} shows the materialized hypercube corresponding to the hypercube shown in Figure \ref{fig-default}.

\begin{figure*}
\centering
\caption{A materialized hypercube. Each row corresponds to one cell. The last column contains the value of the cell, other columns correspond to the dimensions. Default values are automatically inserted, so that this is a highly structured data cube.}
\vspace{3mm}
\label{fig-materialized}
\begin{tabular}{|c|c|c|c|c||c|}
\hline
Concept & Period & Entity & Unit & Region & Value \\
\hline
Assets & Sept. 30th, 2012 & Visto & USD & United States & 3,000,000,000 \\
Assets & Sept. 30th, 2012 & Visto & USD & [World] & 4,000,000,000 \\
Assets & Sept. 30th, 2012 & Championcard & USD & United States & 6,000,000,000 \\
Assets & Sept. 30th, 2012 & Championcard & USD & [World] & 8,000,000,000 \\
Assets & Sept. 30th, 2012 & American Rapid & USD & United States & 5,000,000,000 \\
Assets & Sept. 30th, 2012 & American Rapid & USD & [World] & 9,000,000,000 \\
Equity & Sept. 30th, 2012 & Visto & USD & United States & 2,000,000,000 \\
Equity & Sept. 30th, 2012 & Visto &USD &  [World] & 3,000,000,000 \\
Equity & Sept. 30th, 2012 & Championcard & USD & United States & 4,000,000,000 \\
Equity & Sept. 30th, 2012 & Championcard & USD & [World] & 5,000,000,000 \\
Equity & Sept. 30th, 2012 & American Rapid & USD & United States & 3,000,000,000 \\
Equity & Sept. 30th, 2012 & American Rapid & USD & [World] & 6,000,000,000 \\
Liabilities & Sept. 30th, 2012 & Visto & USD & United States & 1,000,000,000 \\
Liabilities & Sept. 30th, 2012 & Visto & USD & [World] & 1,000,000,000 \\
Liabilities & Sept. 30th, 2012 & Championcard & USD & United States & 2,000,000,000 \\
Liabilities & Sept. 30th, 2012 & Championcard & USD & [World] & 3,000,000,000 \\
Liabilities & Sept. 30th, 2012 & American Rapid & USD & United States & 2,000,000,000 \\
Liabilities & Sept. 30th, 2012 & American Rapid & USD & [World] & 3,000,000,000 \\
\hline
\end{tabular}
\end{figure*}

\subsection{Spreadsheet views}

A hypercube can be materialized into a table as shown in the former section. With the state of matter analogy, it can be seen as the solid version of a (very small) subpart of the gas of cells.

From a business viewpoint, tables are very useful because they can be understood without IT knowledge. However, a materialized hypercube displays the multidimensional data under a very raw form. This raw form is actually very common though, so that mainstream spreadsheet software provide a feature that allows interacting with multidimensional data with a better UI. This feature is often called \emph{pivot table} and flattens the data to a two-dimensional sheet.

The dimensions are partitioned amongst:
\begin{description}
\item [Slicers] All the data that does not match the slicers is discarded.
\item [Dicers] They specify, for each row and column, what dimensional constraints the data at their intersection must fulfill.
\item [Values (potentially aggregated)] They specify, for each cell, which property is displayed as well as, if there are several values, how to aggregate them (sum, count, max, min, average, etc).
\end{description}

This functionality is straight-forward to implement on top of a cell store, because the raw data is in exactly the same form. The XBRL specifications also contain a feature called \emph{table link base} that standardizes how to specify such spreadsheet views.

Figure \ref{fig-spreadsheet} shows an example of how the materialized hypercube on Figure \ref{fig-materialized} can be displayed in this more business-oriented manner.

Concretely, the construction of the view can be pushed to the server or the cell store itself:

\begin{itemize}
\item given a hypercube (and possibly the cells it contains, queried from the store), a spreadsheet can be smartly generated. Slicers are taken from all dimensions that only have a single value across the hypercube, concepts can be assigned to the rows and the remaining dimensions to the columns.
\item given a spreadsheet definition (say, table link base), a hypercube can be generated in order to obtain all the relevant cells from the underlying cell store.
\end{itemize}

It is also still possible for a business user to obtain the materialized hypercube from the cell store, and to import it as-is in their spreadsheet software.

As is commonly done in spreadsheets, business users can drag and drop dimensions across the different categories to fine tune their view over the data. Spreadsheet views are not only convenient to read, but also to write data back to the cell store, cell by cell. Compared to the cell gas and the solid materialized hypercube, the spreadsheet view is comparable to a metal that gets melted before the blacksmith can shape it at will.

Hence, because cell stores can use all the experience accumulated over several decades on pivot tables from the spreadsheet industry, they offer a powerful and business friendly interface, shielding users from the underlying dimensional complexity.

To a business user, working with a cell store feels like working on a spreadsheet, except that:

\begin{itemize}
\item the size of the data is orders of magnitude bigger than a spreadsheet file;
\item the data lies on a server and is shared across a department or a company;
\item the latest database technologies are leveraged under the hood to scale up and out, without the need to go through the IT department for each change in the business taxonomy.
\end{itemize}

\begin{figure*}
\centering
\caption{A spreadsheet view over a hypercube, for viewing and editing data without IT knowledge. The display style used here is done in the spirit of XBRL table link bases. In this case, concepts are put on rows and the other dimensions on filters or on columns. The spreadsheet front end can support drag-and-drop, allowing the user to interactively rearrange rows, columns and filters. Note how default values are handled with L-shape cells.}
\label{fig-spreadsheet}
\vspace{3mm}
\begin{tabular}{|l||c|c|c|c|c|c|}
\hline
\textbf{Unit} & \multicolumn{6}{|l|}{USD} \\
\hline
\textbf{Period} & \multicolumn{6}{|l|}{Sept. 30th, 2012 } \\
\hline
\hline
\multirow{4}{*}{Line items} & \multicolumn{6}{|c|}{\textbf{Entity}} \\
\cline{2-7}

& \multicolumn{2}{|c|}{Visto} & \multicolumn{2}{|c|}{Championcard} & \multicolumn{2}{|c|}{American Rapid} \\
\cline{2-7}

& \multicolumn{2}{|c|}{\textbf{Region}} & \multicolumn{2}{|c|}{\textbf{Region}} & \multicolumn{2}{|c|}{\textbf{Region}} \\
\cline{2-2}\cline{4-4}\cline{6-6}

& United States & & United States &  & United States & \\
\hline
\hline

Assets & 3,000,000,000 & 4,000,000,000 & 6,000,000,000 & 8,000,000,000 & 5,000,000,000 & 9,000,000,000 \\
\hline

Equity & 2,000,000,000 & 3,000,000,000 & 4,000,000,000 & 5,000,000,000 & 3,000,000,000 & 6,000,000,000 \\
\hline

Liabilities & 1,000,000,000 & 1,000,000,000 & 2,000,000,000 & 3,000,000,000 & 2,000,000,000 & 3,000,000,000 \\
\hline
\end{tabular}
\end{figure*}

\subsection{Maps}

When many people define their own taxonomy, this often ends up in redundant terminology. For example, someone might use the term Equity and somebody else Capital. When either querying cells with a hypercube, or loading cells into a spreadsheet, a mapping can be applied so that this redundant terminology is transparent. This way, when a user asks for Equity, (i) she will also get the cells having the concept Capital, (ii) and it will be transparent to her because the Capital concept is overridden with the expected value Equity.

\subsection{Rules}

One of the reasons spreadsheets are very popular is that formulas can be entered into cells to automatically compute values.

Cell stores support an equivalent capability called \emph{rules}. Like maps, rules are executed in a transparent way during a hypercube query, or when a spreadsheet is requested.

From a high-level perspective, cell stores support two kinds of rules:

\begin{description}
\item[Imputation] rules compute a value for a missing cell (i.e., dimensional coordinates against which no value was reported). When generated, this cell comes along with an audit trail that indicates how the value was computed, and from which other cells.
\item[Validation] rules check that the value for a given cell is consistent with the values reported in other cells (often neighbors in the spreadsheet view). A validation rule typically results in a green tick or a red cross in the corresponding cell on the spreadsheet view.
\end{description}

Rules can be defined according to several metapatterns, as defined by Charles Hoffman. It is most intuitive to think about them having in mind the spreadsheet view.

\begin{description}
\item [Roll Up] Several cells with different concepts (but with the exact same other dimensions) are aggregated (often with a sum) into a roll up value. This corresponds to summing across a column or row in Excel.
\item [Roll Forward] A value for a new instant in time is deduced from the equivalent value at a former time, as well as from the delta value on the corresponding time interval.
\item [Compound Fact] This is the same as a roll up, except that instead of the concept varying, the aggregation is computed against a different dimension.
\item [Adjustment] This is similar to a roll forward, except that the time correspond to transaction time, not valid time, and the delta corresponds to a correction or an amendment.
\item [Variance] This is similar to a compound fact, except that the dimension used has the semantics of two different scenarios.
\item [Complex Computation] This is a generalized roll up.
\item [Grid] This metapattern involves the conjunction of two other metapatterns, for example, in the spreadsheet view, a roll up on the rows and a compound fact on the columns.
\end{description}

To facilitate the definition of rules, concepts and dimension values are organized in hierarchies. For example, a roll up is typically a parent's being the sum of its children. Arborescent formats such as JSON or XML cover this need well.

\section{Canonical mapping to the relational model}
\label{section-relational-mapping}
There is a direct, two-way canonical mapping between hypercubes and relational tables.

A materialized hypercube can be converted to a relational table with equivalent semantics by removing the Concept column, and replacing the Value column with one column for each concept, as shown on Figure \ref{fig-relationalized}. The set of all attributes corresponding to the dimension columns acts as a primary key to the table.

\begin{figure*}
\caption{A relational table corresponding to a hypercube. The \emph{concept} dimension is handled in a special way: all cells that have the same dimensions, but \emph{concept}, are grouped in a business object, and displayed in the same row.}
\centering
\label{fig-relationalized}
\vspace{3mm}
\begin{tabular}{|c|c|c|c||c|c|c|}
\hline
Period & Entity & Unit & Region & Assets & Equity & Liabilities \\
\hline
Sept. 30th, 2012 & Visto & USD & United States & 3,000,000,000 & 2,000,000,000 & 1,000,000,000 \\
Sept. 30th, 2012 & Visto & USD & [World] & 4,000,000,000 & 3,000,000,000 & 1,000,000,000 \\

Sept. 30th, 2012 & Championcard & USD & United States & 6,000,000,000 & 4,000,000,000 & 2,000,000,000 \\

Sept. 30th, 2012 & Championcard & USD & [World] & 8,000,000,000 & 5,000,000,000 & 3,000,000,000 \\

Sept. 30th, 2012 & American Rapid & USD & United States & 5,000,000,000 & 3,000,000,000 & 2,000,000,000 \\

Sept. 30th, 2012 & American Rapid & USD & [World] & 9,000,000,000 & 6,000,000,000 & 3,000,000,000 \\

\hline
\end{tabular}
\end{figure*}

Conversely, any relational table can be converted to a cell gas and its corresponding hypercube as follows: each attribute in the primary key is converted to a dimension. A cell is
then created for each row and for each value on that row that is not a primary key. This cell is associated with the dimensions values corresponding to the primary keys on the same row, plus the \emph{concept} dimension associated with the name of the attribute corresponding to the column.

The consequence of this is that an entire relational database with multiple tables, or even several relational databases, can be converted into a single cell store. Likewise, relational views can be built dynamically on top of a cell store.

A hypercube query corresponds to both a relational algebra projection and selection: a projection is nothing else than a selection done on the \emph{concept} dimension.

\section{Standardized Data Interchange: XBRL}
\label{section-xbrl-standard}
Many ideas behind the cell store paradigm originate from the XBRL standard \cite{XBRL}. There are three main reasons for this:

\begin{itemize}
\item The XBRL standard was designed by a team who is aware of the needs of business users, and of the challenges of business reporting.
\item It is important that the data stored in a cell store is not locked in this cell store, i.e., that it can be exported in such a way that other users, even not cell store users, can understand it and use it without ETL efforts. The XBRL standard makes sure that this is so: cell stores can import XBRL data, and export their content into the XBRL format.
\item Cell stores were designed with the goal of providing efficient storage and retrieving capabilities for XBRL data. They provide an abstract data model on top of XBRL that is a viable and efficient alternative to other implementations, such as storing each XBRL hypercube in ROLAP, or such as importing raw XBRL filings into an XML database.
\end{itemize}

XBRL is complex and involves many different specifications.

XBRL, on the physical level, uses XML technologies: filings are reported with a (flat) XML format, and metadata (called taxonomies) using XML Schema and XLink.

The counterpart of a cell is called a fact. A numeric fact may also be stamped with information on the precision or the number of decimals.

In XBRL, dimensions are called aspects. There are three ``builtin'' aspects in addition to \emph{concept}: \emph{period}, \emph{entity} and \emph{unit}, and taxonomies may define more dimensions.

Taxonomies define concepts, hypercubes, dimensions, dimension values (members), etc. They can be shared at any level (i.e., reporting authority, company, department, etc) and extended at will.

XBRL Link bases provide metadata information in the form of ``networks'', among which:

\begin{description}
\item[Definition networks] They allow, among others, building hypercubes and specifying which concepts are bound to which hypercubes, which hypercubes have which dimensions, and which dimensions have which values. Dimensions may either have a typed value space, or be an explicitly enumerated set. 
\item[Presentation networks] They allow the hierarchization of concepts and dimension values in a spreadsheet view (i.e., a table link base in XBRL). For example, a balance sheet may be divided into an Assets hierarchy and an EquityAndLiabilities hierarchy. A presentation network can also contain abstract concepts, i.e., they are only here to organize and partition other concepts. 
\item[Table link bases] They define how a spreadsheet view looks like, i.e., which are the slicers, the dicers, how the dicers are organized in rows and columns, etc.
\item[Calculation networks] They are the simplest kind of roll up rules.
\item[Label networks] They associate business-friendly labels to concepts, dimensions and dimension values, because the latter are often stored in a very raw form that is not palatable to non IT-savvy users.
\item[Formula networks] They define rules to automatically impute or validate fact values.
\end{description}

\section{Implementation}
We now give details on the existing implementation on top of MongoDB (NoLAP), as well as hints on how cell stores could also be implemented on top of other kinds of stores (in particular column stores).
\label{section-implementation}

\subsection{On top of a document store: NoLAP}

The first cell store was implemented on top of a document store (MongoDB \cite{MongoDB}), entirely with the JSONiq language \cite{JSONiq}. The in-memory processing is performed by the Zorba engine \cite{Zorba}. The ETL was made with an existing XBRL processor, in Java. It contains fiscal information reported by public US companies to the SEC. The data is available publicly \cite{SECXBRL.info} for Dow 30 companies. Hypercube queries or spreadsheet queries can be made via a REST API.

From a document store metadata perspective, the implementation is very simple, as only two collections are used:

\begin{figure}
\caption{A cell represented as a JSON object (fact). This is a simplified view, as additional fields may be added in order to optimize queries. The value field is typed and the types are mapped to BSON.}
\label{fig-fact}
\begin{lstlisting}
{
  "Aspects" : {
    "Concept" : "Assets",
    "Period" : "2012-09-30",
    "Entity" : "Visto",
    "Unit" : "USD"
  }
  "Value" : 4000000000
}
\end{lstlisting}
\end{figure}

\begin{description}
\item[facts] This is where the data lies. The SEC repository \cite{SECXBRL.info} contains the order of magnitude of one hundred million facts (200 GB). Each fact is a JSON object as depicted on Figure \ref{fig-fact}. Several indexes on the fields used most (concept, entity) make sure hypercube queries are efficient. Hypercube queries can directly be translated to MongoDB queries, and hence almost completely pushed to the server backend.
\item[components] The metadata is stored here (ca. 100 GB). Each component contains a hierarchy of concepts, a couple of hypercubes, a spreadsheet definition, business rules, concept metadata such as labels in various languages, and documentation. Given a component, data cubes or spreadsheet views can be built.
\end{description}

Some additional data such as XBRL filings and filer information, mostly structured, is stored in further collections. However, in view of the relational mapping depicted in Section \ref{section-relational-mapping}, it is planned to also push this data to the cell store itself.

Another collection (concepts) is used in order to optimize querying for concepts, including full text search, and finding out which components they appear in.

\subsection{On top of a column store (e.g., Cassandra)}

From a theoretical viewpoint, a cell store could fit in a sparse Cassandra table, but this would not scale up well: this would require as many primary keys as dimensions, as well as the materialization of default dimension values to special primary key values.

Rather, dimensions could be set up as non-primary-key columns, using a UUID primary key instead. Secondary indices on the dimensional columns ensure efficient hypercube retrieval. In order to take advantage of the flexibility of Cassandra with respect to columns, the \emph{concept} dimension could be handled separately, with all cells corresponding to the same business object (that is, all dimensions but \emph{concept} have the same values) on the same row. This corresponds to the relational mapping mentioned in Section \ref{section-relational-mapping}.

\subsection{On top of a key-value store}

The data in a cell store could be stored in a key-value store, possibly in an optimized format for retrieval and for saving space. However, document stores are more suitable for the storage of metadata, as tree structures are still quite useful for modeling business taxonomies.

\subsection{On top of a graph database}

A cell store could be stored in a graph database, by splitting each cell into several triples: the subject is the cell, it has one predicate for each dimension leading to the dimension value (as an object), and a predicate leading to the cell value. However, this would both lead to an increased number of ``atoms'' (on the order of magnitude of ten times more), and to inefficient retrieval, as each cell must be reassembled from the triples.

\section{Performance}
\label{section-performance}
Measurements were performed on top of the first cell store repository, secxbrl.info.

It contains 200 GB of data (75 millions of cells).

Hypercube queries are done via a REST API, implemented in JSONiq and executed with the underlying Zorba engine. The computation is done on Amazon EC2 machines, and the data is hosted by compose.io, also on Amazon machines in the same region (US East).

It can be seen that retrieval is (by one or two orders of magnitude) less efficient than lower level NoSQL stores (because it is run on top of a NoSQL store, with additional processing machinery). The end goal of this first implementation is to show feasibility (proof of concept), and to show that hypercube queries, maps, rules and spreadsheet queries are performed in a time acceptable by a human, that is, not more than a couple of seconds. This goal is achieved for these simple use cases, and for a repository size above the lower end of typical benchmarks (i.e., 100 GB). Future cell store implementations (for example, a native cell store) are likely to improve this performance: as a rule of thumb, we hope to show that it should be feasible to achieve hypercube query performance close to that of a filtering query in a document store.

\begin{figure*}
\caption{Typical execution times. These were obtained on the proof-of-concept implementation on top of MongoDB, on a repository with 75 millions of cells. The queries were executed through a hypercube-building REST API, and an average was taken on 20 executions). All times are below the threshold acceptable for human interaction.}
\centering
\label{fig-measurements}
\vspace{3mm}
\begin{tabular}{|l|l|l|}
\hline
Type of query & Number of cells & Time \\
\hline
\hline
Point query & 1 & 200 ms \\
\hline
Query across a row&30& 300 ms \\
\hline
Query across two dimensions&124& 400 ms \\
\hline
Query of all (raw) cells in a component &82& 250 ms \\
\hline
Query of all cells in a component, including mapping, rule execution and validation& 96& 1500 ms \\
\hline
Building a spreadsheet out of a component, including mapping, rule execution and validation & 96 & 1900 ms \\
\hline
\end{tabular}
\end{figure*}

\section{Future Work}

A future version of this paper will contain more elaborate comments on performance and more measurements. Also, we are considering populating a database from the TPC-H benchmark with the relational mapping, and executing TPC-H queries on top, in order to compare with other implementations.

The current cell store implementation focuses on hypercube queries, maps, rules and spreadsheet queries. In the future, it will be desirable to integrate cell stores tightly with MDX and SQL, and investigate how well cell stores scale up for more sophisticated operations such as joins. In other words, MDX queries can be translated to hypercube queries, and SQL queries can be translated to hypercube queries and some additional JSONiq machinery using the relational mapping.

Also, we will continue to work on improving performance. In a farther future, we aim at proving that our assumption that a native cell store implementation (rather than the two-layer approached taken for our first implementation) will deliver significantly better performance.

\section{Conclusion}

Cell stores leverage the latest database technologies, but completely give control over their data to business users. The data is stored at the cell level, in a single, big collection (gas of cells) and can be clustered, replicated, and retrieved efficiently. Cells can be assembled into data cubes with hypercube queries, and assembled into spreadsheet views, with which business users can interact (read, write) with the data. Business users can define their own taxonomies, schemas, maps, rules without any interaction with the IT department.

\section{Acknowledgements}

This is joint work. The implementation of the first cell store on top of a document store has been made as a team effort by Matthias Brantner, William Candillon, Federico Cavalieri (also to be thanked for performing several rounds of proof reading), Dennis Knochenwefel, Alexander Kreutz and myself. We received a lot of very useful advice from Charles Hoffman.

\bibliographystyle{abbrv}
\bibliography{cellstores}

\end{document}